\documentclass[%
 reprint,
 amsmath,amssymb,
 aps,
prb,
 superscriptaddress
]{revtex4-2}

\usepackage{graphicx}
\usepackage{dcolumn}
\usepackage{bm}
\usepackage{braket}
\usepackage{color}

\begin{document}

\preprint{APS/123-QED}

\title{Scaling rule in critical non-Hermitian skin effect}

\author{Kazuki Yokomizo}
\affiliation{Condensed Matter Theory Laboratory, RIKEN, 2-1 Hirosawa, Wako, Saitama, 351-0198, Japan}
\author{Shuichi Murakami}%
\affiliation{Department of Physics, Tokyo Institute of Technology, 2-12-1 Ookayama, Meguro-ku, Tokyo, 152-8551, Japan}
\affiliation{TIES, Tokyo Institute of Technology, 2-12-1 Ookayama, Meguro-ku, Tokyo, 152-8551, Japan}%




%
\begin{abstract}
Non-Hermitian systems show a non-Hermitian skin effect, where the bulk states are localized at a boundary of the systems with open boundary conditions. In this paper, we study the dependence of the localization length of the eigenstates on a system size in a specific non-Hermitian model with a critical non-Hermitian skin effect, where the energy spectrum undergoes discontinuous transition in the thermodynamic limit. We analytically show that the eigenstates exhibit remarkable localization, known as scale-free localization, where the localization length is proportional to a system size. Our result gives theoretical support for the scale-free localization, which has been proposed only numerically in previous works.
\end{abstract}
\pacs{Valid PACS appear here}
\maketitle
%
%

\section{\label{sec1}Introduction}
Nonequilibrium and open systems are effectively described in terms of non-Hermitian Hamiltonians. Thus far, non-Hermitian systems have been attracting much attention to study remarkable phenomena, such as unidirectional transmission~\cite{Guo2009,Feng2013} and retroreflection~\cite{Wang2019}. Among theoretical and experimental studies on non-Hermitian systems, a non-Hermitian skin effect plays an important role in condensed matter physics. In the non-Hermitian skin effect, eigenstates of the bulk are localized at the end of a system with an open boundary condition~\cite{Yao2018,Yoshida2020,Okuma2020,Zhang2020,Kawabata2020v2,Okugawa2020,Fu2021,Okugawa2021}. Furthermore, this effect gives rise to the difference between energy spectra in an open chain and those in a periodic chain. These phenomena have been experimentally observed in various physical systems~\cite{Brandenbourger2019,Gou2020,Xiao2020,Weidemann2020,Hofmann2020,Helbig2020,Ghatak2020}. It is significant to investigate some phenomena induced by the non-Hermitian skin effect because this effect can lead to non-Hermitian phenomena~\cite{Song2019,Longhi2020,Yu2020,Yu2020v2,Lee2020,Yi2020,Liu2020,Flebus2020,Xue2021}.

Additionally, in recent years, non-Hermitian systems have been much investigated from various perspectives. For example, Ref.~\cite{Li2020} theoretically proposed a class of criticality. The authors showed a critical non-Hermitian skin effect, where an energy spectrum and localization of eigenstates discontinuously jump across a critical point. We note that, since the critical non-Hermitian skin effect occurs in the thermodynamic limit, this effect can be systematically understood in terms of a non-Bloch band theory~\cite{Yao2018,Yokomizo2019,Kawabata2020,Yokomizo2020,Yokomizo2020v2,Yi2020,Yang2020,Yokomizo2021}. This is because, in non-Hermitian systems with open boundary conditions, the non-Bloch band theory can calculate continuum energy bands in the limit of a large system size. On the other hand, Ref.~\cite{Li2020} also proposed that when a system with the critical non-Hermitian skin effect has a finite system size, the eigenstates exhibit remarkable localization behavior. It is called {\it scale-free localization}, where the localization length is proportional to the system size. Importantly, the non-Bloch band theory cannot study the scale-free localization because this localization occurs only in a finite open chain. Therefore, the nature of the scale-free localization has not been revealed yet.

Thus far, finite non-Hermitian systems have been investigated only by a numerical calculation. Hence, it is necessary to study non-Hermitian systems with a finite system size in more detail to reveal the nature of non-Hermitian phenomena unique to a finite system, such as the scale-free localization. Furthermore analytical studies of such systems will give us useful insights into experimental studies on the non-Hermitian skin effect.

In this paper, we analyze a scaling rule of the localization length of the eigenstates in a ladder model. As a result, we find remarkable phenomena unique to a finite open non-Hermitian system as follows. When the system size is sufficiently small, the energy eigenvalues and the eigenstates are insensitive to the change of the system size. On the other hand, when the system size is larger than a critical value, the localization length changes in proportion to the system size, and the energy eigenvalues also change as the system size changes. Then in our analysis, it is shown that these eigenstates exhibit the scale-free localization. Furthermore, we can derive a formula of the critical value of system size.

This paper is organized as follows. In Sec.~\ref{sec2}, we introduce the ladder model and describe the critical non-Hermitian skin effect. Furthermore we explain the concept of our study on a finite open chain. In Sec.~\ref{sec3}, we derive an analytical expression of the localization length and the critical system size in the ladder model. Then we confirm that our result can precisely predict the critical value compared with the result of Ref.~\cite{Okuma2019}. Finally, we summarize the results of this paper and comment on extension to general non-Hermitian systems.

%
%

\section{\label{sec2}Critical non-Hermitian skin effect}
\begin{figure}[]
\includegraphics[width=8.5cm]{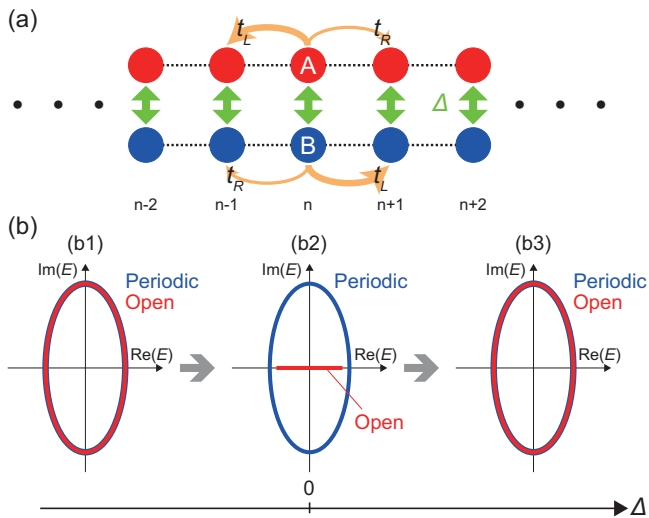}
\caption{\label{fig1}(a) Ladder model in an infinite chain. The red and blue circles represent the sites in sublattices ${\rm A}$ and ${\rm B}$, respectively. We set the hopping amplitude to be $t_L>t_R>0$. The parameter $\Delta$ represents the coupling between two Hatano-Nelson models. (b) Energy spectra of the ladder model in an infinite chain with open (red) and periodic (blue) boundary conditions. The energy spectra in an open chain discontinuously jump across the critical point $\Delta=0$.}
\end{figure}
In this section, we briefly review the critical non-Hermitian skin effect. First, we introduce a tight-binding model on a one-dimensional ladder, as shown in Fig.~\ref{fig1}(a). It consists of two one-dimensional chains with lattice sites ${\rm A}$ and ${\rm B}$. The real-space Hamiltonian of this system is given by
\begin{eqnarray}
H&=&\sum_n\left(t_Lc_{n,{\rm A}}^\dag c_{n+1,{\rm A}}+t_Rc_{n+1,{\rm A}}^\dag c_{n,{\rm A}}+\Delta c_{n,{\rm A}}^\dag c_{n,{\rm B}}\right. \nonumber\\
&&\left.+t_Rc_{n,{\rm B}}^\dag c_{n+1,{\rm B}}+t_Lc_{n+1,{\rm B}}^\dag c_{n,{\rm B}}+\Delta c_{n,{\rm B}}^\dag c_{n,{\rm A}}\right),
\label{eq1}
\end{eqnarray}
where $c_{n,\mu}~(\mu={\rm A},{\rm B})$ represents a fermionic annihilation operator at the $n$th unit cell in the $\mu$ sublattice. For simplicity, the parameters $t_L,t_R$, and $\Delta$ are set to be real, and we assume $t_L>t_R>0$. Below, we explain the behavior of the ladder model in the limit of a large system size.

When $\Delta=0$, the system is decoupled into two Hatano-Nelson models~\cite{Hatano1996}. In this case, the two Hatano-Nelson models exhibit the non-Hermitian skin effect, and the bulk eigenstates in one chain and those in the other chain are localized at the opposite ends of the chain. Furthermore, the energy spectrum in an open chain takes different values from that in a periodic chain, as shown in Fig.~\ref{fig1}(b2). On the other hand, by coupling the two chains with $\Delta\neq0$, the non-Hermitian skin effect disappears because the two skin modes are coupled to each other. Then in contrast to the case of $\Delta=0$, the energy spectrum in an open chain coincides with that in a periodic chain, as shown in Figs.~\ref{fig1}(b1) and \ref{fig1}(b3). Thus, the energy spectrum in the ladder model with open boundary conditions discontinuously jumps across the critical point $\Delta=0$. In Ref.~\cite{Li2020}, this discontinuous phase transition was explained in terms of the critical non-Hermitian skin effect.

\begin{figure}[]
\includegraphics[width=8.5cm]{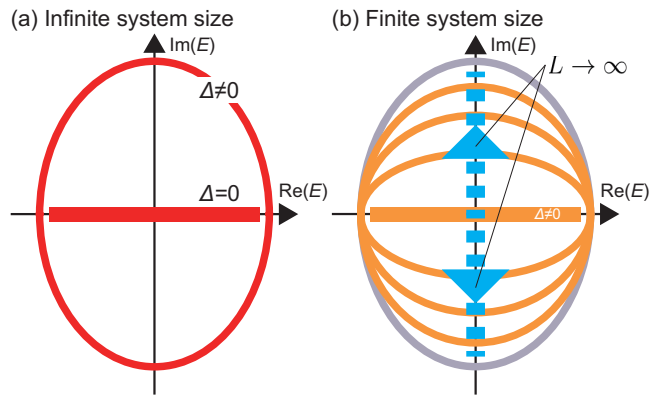}
\caption{\label{fig2}Energy eigenvalues of the ladder model (a) in an infinite open chain and (b) in a finite open chain. (a) In the limit of a large system size, by adding the coupling parameter $\Delta$, the energy spectrum discontinuously jumps. (b) When $\Delta\neq0$, the energy eigenvalues in a finite open chain (orange) gradually approach those in a periodic chain (gray) as the system size increases. This change is represented by the cyan arrows.}
\end{figure}

In contrast, when the system size $L$ is finite, the energy eigenvalues behave continuously as a function of $\Delta$. Depending on the order of the limits $L\rightarrow\infty$ and $\Delta\rightarrow0$, the behavior of the system is critically different~\cite{Okuma2019}. Namely, after taking the thermodynamic limit, the infinitesimal coupling discontinuously changes the energy spectrum [Fig.~\ref{fig2}(a)]. On the other hand, under $0<\Delta\ll{\cal O}\left(t_L,t_R\right)$, an increase of the system size amplifies the  coupling strength. As a result, the energy eigenvalues in an open chain gradually change as the system size increases, and finally, they approach those in a periodic chain in the limit of $L\rightarrow\infty$ [Fig.~\ref{fig2}(b)]. In this paper, we theoretically investigate the energy eigenvalues and the corresponding eigenstates of the ladder model as the system size changes.

%
%

\section{\label{sec3}Scaling rule}
In this section, first, we show a way to get the energy eigenvalues in the ladder model. Next, we obtain system size dependence of the solutions of the characteristic equation for $\beta={\rm e}^{ik}$, where $k$ corresponds to the Bloch wave number. Accordingly, it is shown that the eigenstates exhibit the scale-free localization in the ladder model with a finite system size. We also reveal dependence of the corresponding energy eigenvalues on the system size. In the following, we assume $0<\Delta\ll{\cal O}\left(t_L,t_R\right)$ for simplicity.

%
%

\subsection{\label{sec3-1}Periodic and open chains}
In the ladder model with a periodic boundary condition, the energy eigenvalues $E_{\rm PBC}$ are obtained from the Bloch Hamiltonian:
\begin{eqnarray}
H\left(k\right)=\left( \begin{array}{cc}
t_L{\rm e}^{ik}+t_R{\rm e}^{-ik} & \Delta                           \vspace{5pt}\\
\Delta                           & t_R{\rm e}^{ik}+t_L{\rm e}^{-ik}
\end{array}\right),
\label{eq2}
\end{eqnarray}
as
\begin{eqnarray}
E_{\rm PBC}^\pm\left(k\right)=\left(t_L+t_R\right)\cos k\pm i\sqrt{\left(t_L-t_R\right)^2\sin^2k-\Delta^2}, \nonumber\\
\label{eq3}
\end{eqnarray}
where $k$ is the real Bloch wave number.

On the other hand, in the ladder model with an open boundary condition, the energy eigenvalues are calculated from a real-space Schr\"{o}dinger equation. In the following, let $E_{\rm OBC}$ denote the energy eigenvalues in the open chain. Then we can explicitly write the equation $H\ket{\psi}=E_{\rm OBC}\ket{\psi}$, where $\ket{\psi}=\left(\dots,\psi_{1,{\rm A}},\psi_{1,{\rm B}},\dots,\psi_{n,{\rm A}},\psi_{n,{\rm B}},\dots\right)^{\rm T}$, as
\begin{eqnarray}
\left\{ \begin{array}{l}
t_R\psi_{n-1,{\rm A}}+\Delta\psi_{n,{\rm B}}+t_L\psi_{n+1,{\rm A}}=E_{\rm OBC}\psi_{n,{\rm A}}, \vspace{3pt}\\
t_L\psi_{n-1,{\rm B}}+\Delta\psi_{n,{\rm A}}+t_R\psi_{n+1,{\rm B}}=E_{\rm OBC}\psi_{n,{\rm B}}.
\end{array}\right.
\label{eq4}
\end{eqnarray}
According to the theory of linear difference equations, we can take an ansatz for the eigenstates as a linear combination:
\begin{eqnarray}
\left( \begin{array}{c}
\psi_{n,{\rm A}} \vspace{5pt}\\
\psi_{n,{\rm B}}
\end{array}\right)=\sum_{j=1}^4\left(\beta_j\right)^n\left( \begin{array}{c}
\phi_{\rm A}^{\left(j\right)} \vspace{5pt}\\
\phi_{\rm B}^{\left(j\right)}
\end{array}\right).
\label{eq5}
\end{eqnarray}
Hence Eq.~(\ref{eq4}) can be rewritten as
\begin{eqnarray}
\left( \begin{array}{cc}
t_L\beta+t_R\beta^{-1} & \Delta                 \vspace{5pt}\\
\Delta                 & t_R\beta+t_L\beta^{-1}
\end{array}\right)\left( \begin{array}{c}
\phi_{\rm A} \vspace{5pt}\\
\phi_{\rm B}
\end{array}\right)=E_{\rm OBC}\left( \begin{array}{c}
\phi_{\rm A} \vspace{5pt}\\
\phi_{\rm B}
\end{array}\right), \nonumber\\
\label{eq6}
\end{eqnarray}
with $\beta=\beta_j$, and $\phi_\alpha=\phi_\alpha^{\left(j\right)}~(\alpha={\rm A},{\rm B})$. Here, we can obtain the characteristic equation as the condition that the coefficients $\phi_{\rm A}$ and $\phi_{\rm B}$ take nonzero values, explicitly written as
\begin{eqnarray}
\beta^2&-&\left(\frac{1}{t_L}+\frac{1}{t_R}\right)E_{\rm OBC}\beta+\frac{1}{t_Lt_R}\left(t_L^2+t_R^2+E_{\rm OBC}^2-\Delta^2\right) \nonumber\\
&-&\left(\frac{1}{t_L}+\frac{1}{t_R}\right)E_{\rm OBC}\beta^{-1}+\beta^{-2}=0.
\label{eq7}
\end{eqnarray}

Now we assume that the system has $L$ unit cells, and we impose the open boundary condition $\psi_{0,\mu}=\psi_{L+1,\mu}=0~(\mu={\rm A},{\rm B})$ to this system. Then from Eq.~(\ref{eq4}), we can derive the boundary equation for the solutions of the characteristic equation (\ref{eq7}) as
\begin{eqnarray}
&&X_{1,4}X_{2,3}\left[\left(\beta_1\beta_4\right)^{L+1}+\left(\beta_2\beta_3\right)^{L+1}\right] \nonumber\\
&&-X_{1,3}X_{2,4}\left[\left(\beta_1\beta_3\right)^{L+1}+\left(\beta_2\beta_4\right)^{L+1}\right] \nonumber\\
&&+X_{1,2}X_{3,4}\left[\left(\beta_1\beta_2\right)^{L+1}+\left(\beta_3\beta_4\right)^{L+1}\right]=0,
\label{eq8}
\end{eqnarray}
where $\beta_j~(j=1,\dots,4)$ satisfy $\left|\beta_1\right|\leq\dots\leq\left|\beta_4\right|$, and $X_{i,j}~(i,j=1,\dots,4)$ are defined as
\begin{equation}
X_{i,j}=t_L\left(\beta_j-\beta_i\right)+t_R\left(\beta_j^{-1}-\beta_i^{-1}\right),~\left(i,j=1,\dots,4\right).
\label{eq9}
\end{equation}
We note that the solutions satisfy
\begin{equation}
\beta_1=\frac{1}{\beta_4},~\beta_2=\frac{1}{\beta_3}
\label{eq10}
\end{equation}
because Eq.~(\ref{eq7}) is a reciprocal equation for $\beta$. In Appendix~\ref{secA}, we give the derivation of Eq.~(\ref{eq8}) in the ladder model. Since the characteristic equation (\ref{eq7}) and the boundary equation (\ref{eq8}) determine a set of solutions for $\left(\beta,E\right)$, the energy eigenvalues in the finite open chain can be calculated by combining these equations.

We note that, in the ladder model with $\Delta\neq0$, the energy spectrum in an open chain is equal to that in a periodic chain in the thermodynamic limit. This can be understood in terms of the non-Bloch band theory~\cite{Yokomizo2019,Yokomizo2020}. In Appendix~\ref{secB}, we show a way to get the energy spectrum in an infinite open chain.

%
%

\subsection{\label{sec3-2}Scale-free localization}
Now we focus on the boundary equation (\ref{eq8}). In the following, we assume that the system size is large. In this case, we can approximate Eq.~(\ref{eq8}) by neglecting the terms other than the two dominant terms $-X_{1,3}X_{2,4}\left(\beta_2\beta_4\right)^{L+1}$ and $-X_{1,2}X_{3,4}\left(\beta_3\beta_4\right)^{L+1}$ on the left-hand side. By substituting the solutions of the characteristic equation (\ref{eq7}) into this approximated boundary equation, we can rewrite Eq.~(\ref{eq8}) as
\begin{equation}
\left(\beta_2\right)^{2L+2}\simeq\frac{\left(t_L+t_R\right)^2}{\left(t_L-t_R\right)^2}\frac{E_{\rm OBC}^2-4t_Lt_R}{\left[\left(t_L+t_R\right)^2-E_{\rm OBC}^2\right]^2}\Delta^2
\label{eq11}
\end{equation}
up to the second order of the coupling parameter $\Delta$. We explain a way to derive Eq.~(\ref{eq11}) in Appendix~\ref{secC}.

Next, we estimate the right-hand side in Eq.~(\ref{eq11}). As mentioned in Sec.~\ref{sec2}, in the thermodynamic limit, the energy spectrum in an open chain coincides with that in a periodic chain. Accordingly, in a system with a large system size, the energy eigenvalues in an open chain asymptotically approach those in a periodic chain, as shown in Fig.~\ref{fig2}(b). Hence, we can approximate the energy eigenvalues $E_{\rm OBC}$ by the energy eigenvalues $E_{\rm PBC}^\pm\left(k\right)$ [Eq.~(\ref{eq3})] in a periodic chain. Here, we take $E_{\rm PBC}^+=\left(t_L+t_R\right)\cos k+i\sqrt{\left(t_L-t_R\right)^2\sin^2k-\Delta^2}~(0\leq k\leq\pi/2)$ as $E_{\rm OBC}$. As a result, we can get an analytical expression of the absolute value of $\beta_2$ as
\begin{equation}
\left|\beta_2\right|\simeq\left[\frac{t_L+t_R}{t_L-t_R}f\left(k\right)\Delta\right]^{\frac{1}{L+1}},
\label{eq12}
\end{equation}
where
\begin{equation}
f\left(k\right)=\frac{\sqrt{\left|t_L^2{\rm e}^{2ik}+t_R^2{\rm e}^{-2ik}-2t_Lt_R\right|}}{\left|2\left(t_L^2+t_R^2\right)\sin^2k-i\left(t_L^2-t_R^2\right)\sin2k\right|}.
\label{eq13}
\end{equation}

\begin{figure}
\includegraphics[width=8.5cm]{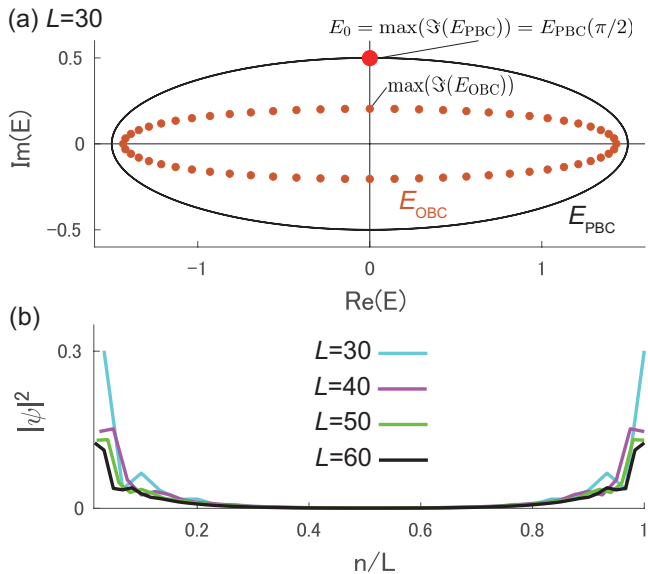}
\caption{\label{fig3}Energy eigenvalues and eigenstates in the ladder model with $t_L=1,t_R=0.5$, and $\Delta=10^{-3}$. (a) Energy eigenvalues in an open chain with the system size $L=30$ ($E_{\rm OBC}$, orange dots) and those in a periodic chain ($E_{\rm PBC}$, black line). The red dot on the black line expresses the maximum value of the imaginary part of $E_{\rm PBC}$ given by $k=\pi/2$ in Eq.~(\ref{eq3}). (b) Spatial distribution of the density of states corresponding to $\max\left[\Im\left(E_{\rm OBC}\right)\right]$ for various system sizes. Their localization lengths divided by the system sizes are equal, which means that the eigenstates have the scale-free localization.}
\end{figure}

Equation~(\ref{eq12}) gives the system size dependence of the solutions of the characteristic equation (\ref{eq7}), which determines localization behavior of the eigenstates. This is because, in a large system, $\beta_2(=1/\beta_3)$ determines the localization length of the eigenstates given by Eq.~(\ref{eq5}). Hence we can obtain the localization length of the eigenstates from Eq.~(\ref{eq12}). For example, when $k=\pi/2$, which corresponds to the energy eigenvalue with a maximum imaginary part [Fig.~\ref{fig3}(a)], Eq.~(\ref{eq12}) can be explicitly written as
\begin{equation}
\left|\beta_2\right|\simeq\left[\frac{\left(t_L+t_R\right)^2}{2\left(t_L-t_R\right)\left(t_L^2+t_R^2\right)}\Delta\right]^{\frac{1}{L+1}}.
\label{eq14}
\end{equation}
Therefore, its localization length, given by $-1/\log\left|\beta_2\right|$, is proportional to the system size, and this localization length becomes larger as the system size increases. This behavior of the eigenstates is called the scale-free localization~\cite{Li2020,Li2021}. In fact, our numerical result of exact diagonalization shows that the localization lengths of the eigenstates divided by the system sizes are the same for various system sizes [Fig.~\ref{fig3}(b)]. Similar scale-free localization appears when $k\neq0$. On the other hand, when $k=0$ in Eq.~(\ref{eq12}), $\beta_2$ approximately becomes $1$ to the same order in Eq.~(\ref{eq14}). This indicates that the corresponding eigenstate is delocalized, and it does not exhibit the scale-free localization.

%
%

\subsection{\label{sec3-3}Energy eigenvalues}
\begin{figure}[]
\includegraphics[width=8.5cm]{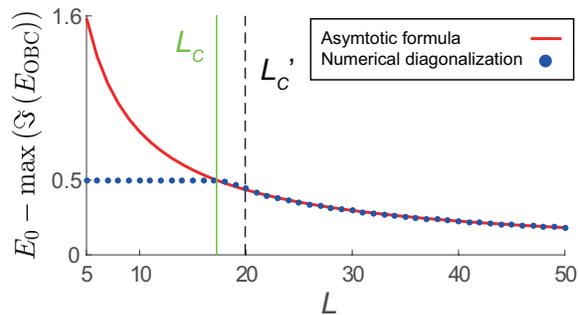}
\caption{\label{fig4}System size dependence of the energy eigenvalues in the ladder model with the open boundary condition. The parameters are set to be $t_L=1,t_R=0.5$, and $\Delta=10^{-3}$. We show the result from the asymptotic formula in Eq.~(\ref{eq15}) in a red line and that from numerical diagonalization in blue dots. $E_0$ is given in Eq.~(\ref{eq16}). $L_c$ is the critical value of our result given in Eq.~(\ref{eq17}), and $L_c^\prime$ is the critical value given in Eq.~(\ref{eq18}) from Ref.~\cite{Okuma2019}. When the system size is smaller than $L_c$, Eq.~(\ref{eq15}) deviates from the result from numerical diagonalization.}
\end{figure}

We can also get the system size dependence of the energy eigenvalues from Eq.~(\ref{eq12}). For simplicity, we focus only on the case with $k=\pi/2$ corresponding to the maximum value of an imaginary part of the energy eigenvalues $\max\left[\Im\left(E_{\rm OBC}\right)\right]$ in the open chain [Fig.~\ref{fig3}(a)]. Now by substituting Eq.~(\ref{eq14}) into Eq.~(\ref{eq7}), $\max\left[\Im\left(E_{\rm OBC}\right)\right]$ can be written as
\begin{equation}
\max\left[\Im\left(E_{\rm OBC}\right)\right]\simeq t_L\left[f\left(\frac{\pi}{2}\right)\Delta\right]^{\frac{1}{L+1}}-t_R\left[f\left(\frac{\pi}{2}\right)\Delta\right]^{-\frac{1}{L+1}}.
\label{eq15}
\end{equation}
As mentioned in Sec.~\ref{sec2}, in the limit of $L\rightarrow\infty$, it is expected to approach the maximum value of an imaginary part of the energy eigenvalues in a periodic chain given by
\begin{equation}
E_0=\sqrt{\left(t_L-t_R\right)^2-\Delta^2}.
\label{eq16}
\end{equation}
In Fig.~\ref{fig4}, we show the value of $E_0-\max\left[\Im\left(E_{\rm OBC}\right)\right]$ both from the asymptotic formula in Eq.~(\ref{eq15}) and from numerical diagonalization. We find that, when the system size is sufficiently large, Eq.~(\ref{eq15}) agrees with the result from numerical diagonalization. On the other hand, when the system size is smaller than a critical value, these results deviate from each other, and in fact, the energy eigenvalues in a finite open chain are independent of the system size. In Sec.~\ref{sec3-4}, we investigate this evolution and the critical system size.

%
%

\subsection{\label{sec3-4}Critical value of system size}
Now we derive an expression of the critical system size as mentioned in Sec.~\ref{sec3-3}. When the system size is sufficiently small, the skin mode of one Hatano-Nelson model in the ladder model is decoupled to the skin mode of the other. We note that the decay of one skin mode is described by the value of $\bar{\beta}=\sqrt{t_R/t_L}$. In fact, its decay length is equal to $-1/\log\bar{\beta}$. Then the energy eigenvalues and the eigenstates are insensitive to the change of the system size. On the other hand, when the system size is sufficiently large, the two skin modes are coupled to each other, and the eigenstates in the ladder model exhibit the scale-free localization. As mentioned in Sec.~\ref{sec2}, the coupling strength between the two skin modes localized at the opposite ends is enhanced exponentially for a large system size. According to Ref.~\cite{Li2020}, we can estimate the scaled coupling as $\bar{\Delta}\sim\bar{\beta}^{-L/2}$. It may look counterintuitive because the skin modes at the left and right ends become faraway from each other as the system size becomes larger.

To understand this counterintuitive behavior, we note that the skin modes form the energy spectra and correspond to not edge modes but bulk modes, although they are localized at either end of the system. Namely, in an open chain, the energy spectra of the skin modes are insensitive to boundary conditions, which indeed means that the skin mode reflects the bulk physics. Thus, while the skin modes of each Hatano-Nelson model are localized at either end, they are sensitive to the bulk physics, and the coupling $\Delta$ makes the non-Hermitian skin effect disappear. Based on these observations, it is natural that the increase of the system size amplifies the effect of the coupling.

Importantly, the scale-free localization competes with the localization of the original skin mode described by $\bar{\beta}$. For example, in the case of $k=\pi/2$, the degree of the delocalization of the eigenstates is obtained from Eq.~(\ref{eq14}) because the more $\left|\beta_2\right|$ approaches $1$, the broader the eigenstates extend over the whole system. In conclusion, the scale-free localization occurs when Eq.~(\ref{eq14}) exceeds $\bar{\beta}=\sqrt{t_R/t_L}$. From the above discussion, we can get the critical system size as
\begin{equation}
L_c=\frac{2}{\log\left(t_R/t_L\right)}\log\left[\frac{\left(t_L+t_R\right)^2\Delta}{2\left(t_L-t_R\right)\left(t_L^2+t_R^2\right)}\right]-1.
\label{eq17}
\end{equation}
In Ref.~\cite{Okuma2019}, the critical value was evaluated in a qualitative manner, given by
\begin{equation}
L_c^\prime=\frac{2\log\Delta}{\log\left(t_R/t_L\right)}.
\label{eq18}
\end{equation}
In Fig.~\ref{fig4}, we show the two critical values for Eqs.~(\ref{eq17}) and (\ref{eq18}) in green and black dashed lines, respectively. We find that our result in Eq.~(\ref{eq18}) can more precisely predict the change from the localization of the original skin mode to the scale-free localization in the ladder model than the result of the previous work in Eq.~(\ref{eq17}).

%
%

\section{\label{sec4}Summary}
In this paper, we study the localization phenomenon in the ladder model. As a manifestation of the critical non-Hermitian skin effect, the eigenstates exhibit the scale-free localization in a finite open chain. Accordingly, the energy eigenvalues gradually change as the system size increases. Furthermore, our analysis can predict the evolution from the localization stemming from the non-Hermitian skin effect to the scale-free localization as the system size becomes larger. It is worth noting that the non-Bloch band theory cannot describe such behavior in the ladder model because the scale-free localization does not occur in the thermodynamic limit.

We discuss universality of the scale-free localization. We focus on non-Hermitian multiband tight-binding systems with open boundary conditions. In this case, we have the characteristic equation $\det\left[{\cal H}\left(\beta\right)-E\right]=0$, and it is an algebraic equation for $\beta$ with an even degree $2M$ in general. Then the behavior of the systems is determined by $\beta_M$ and $\beta_{M+1}$, which are the solutions of the characteristic equation with $\left|\beta_1\right|\leq\dots\leq\left|\beta_{2M}\right|$. Here, we assume that the systems have some symmetries which reverse the direction of a chain, such as inversion symmetry, as is the case in our ladder model. In this case, the characteristic equation is a reciprocal equation, and it follows that $\beta_j=1/\beta_{2M+1-j}~(j=1,\dots,M)$. Hence, we obtain $\beta_M=1/\beta_{M+1}$, and from open boundary conditions, we can get the system size dependence of $\beta_M$ as
\begin{equation}
\left|\beta_M\right|=\left|G\left(\beta_1,\dots,\beta_{2M},E_{\rm OBC}\right)\right|^{\frac{1}{2L}},
\label{eq19}
\end{equation}
In Appendix~\ref{secD}, we explain a way to derive this equation. Equation~(\ref{eq19}) indicates that the eigenstates are localized with the localization length $-1/\log\beta_M$ proportional to a system size. Therefore, the eigenstates exhibit the scale-free localization. In terms of the non-Bloch band theory, this condition is equivalent to that the trajectories of $\beta_M$ and $\beta_{M+1}$, called the {\it generalized Brillouin zone}, become a unit circle. We note that the method of the self-consistent calculation discussed in Sec.~\ref{sec3-2} is useful to analyze behavior of localization in eigenstates.

In this model, the generalized Brillouin zone is given by $\left|\beta_M\right|=\left|\beta_{M+1}\right|=1$ because of the inversion symmetry. From our derivation of the scale-free localization, our theory also applies to a wide class of non-Hermitian systems where $\left|\beta_M\right|=\left|\beta_{M+1}\right|=1$ in the limit of a large system size is guaranteed by some symmetries. Namely, in such systems, if we include a correction due to a finite system size, $\left|\beta_M\right|$ and $\left|\beta_{M+1}\right|$ deviate from 1, and it necessarily leads to the scale-free localization in the same way as in the present case.

Recently, various non-Hermitian systems with the non-Hermitian skin effect were experimentally realized. Therefore, by combining two non-Hermitian systems, we expect the realization of a non-Hermitian system which has eigenstates with the scale-free localization in various physical systems, e.g., an electrical circuit system.

%
%
\begin{acknowledgements}
This paper was supported by JSPS KAKENHI (Grant No.~JP18H03678) and the MEXT Elements Strategy Initiative to Form Core Research Center (TIES) (Grant No.~JPMXP0112101001). K.Y. was also supported by JSPS KAKENHI Grants No.~JP18J22113 and No.~JP21J01409.
\end{acknowledgements}

%
%

\appendix

%
%

\section{\label{secA}Derivation of Eq.~(\ref{eq8})}
In this appendix, we describe the derivation of the boundary equation (\ref{eq8}) in the ladder model with the system size $L$. From the real-space eigenequation in Eq.~(\ref{eq4}) and the open boundary condition $\psi_{0,\mu}=\psi_{L+1,\mu}=0~(\mu={\rm A},{\rm B})$, we can get the equations for the eigenstates in real space as
\begin{eqnarray}
\left\{ \begin{array}{l}
\Delta\psi_{1,{\rm B}}+t_L\psi_{2,{\rm A}}=E_{\rm OBC}\psi_{1,{\rm A}},   \vspace{5pt}\\
\Delta\psi_{1,{\rm A}}+t_R\psi_{2,{\rm B}}=E_{\rm OBC}\psi_{1,{\rm B}},   \vspace{5pt}\\
t_R\psi_{L-1,{\rm A}}+\Delta\psi_{L,{\rm B}}=E_{\rm OBC}\psi_{L,{\rm A}}, \vspace{5pt}\\
t_L\psi_{L-1,{\rm B}}+\Delta\psi_{L,{\rm A}}=E_{\rm OBC}\psi_{L,{\rm B}}.
\end{array}\right.
\label{eqapp1a}
\end{eqnarray}
Now Eq.~(\ref{eqapp1a}) can be rewritten into coupled equations for the coefficients $\phi_\mu^{\left(j\right)}~(\mu={\rm A},{\rm B},j=1,\dots,4)$ by substituting the general solution in Eq.~(\ref{eq5}). Furthermore, by using the bulk eigenequation in Eq.~(\ref{eq6}), we can get the coupled equations including only $\phi_{\rm A}^{\left(j\right)}~(j=1,\dots,4)$. As a result, we can get the condition that $\phi_{\rm A}^{\left(j\right)}~(j=1,\dots,4)$ have nonzero values, written as
\begin{eqnarray}
\left| \begin{array}{cccc}
1                             & 1                             & 1                             & 1                             \vspace{5pt}\\
X_1                           & X_2                           & X_3                           & X_4                           \vspace{5pt}\\
\left(\beta_1\right)^{L+1}    & \left(\beta_2\right)^{L+1}    & \left(\beta_3\right)^{L+1}    & \left(\beta_4\right)^{L+1}    \vspace{5pt}\\
X_1\left(\beta_1\right)^{L+1} & X_2\left(\beta_2\right)^{L+1} & X_3\left(\beta_3\right)^{L+1} & X_4\left(\beta_4\right)^{L+1}
\end{array}\right|=0 \nonumber\\
\label{eqapp2a}
\end{eqnarray}
with $\left|\beta_1\right|\leq\dots\leq\left|\beta_4\right|$. Here, $X_j~(j=1,\dots,4)$ are defined as
\begin{equation}
X_j=E_{\rm OBC}-t_L\beta_j-t_R\beta_j^{-1},~\left(j=1,\dots,4\right).
\label{eqapp3a}
\end{equation}
Finally we can obtain the boundary equation (\ref{eq8}) from Eq.~(\ref{eqapp2a}).

%
%

\section{\label{secB}Non-Bloch band theory}
In this appendix, we briefly review the non-Bloch band theory~\cite{Yokomizo2019}. First, we give the condition for continuum energy bands in non-Hermitian systems with open boundary conditions. Next, we show that, in the ladder model, the continuum energy band in an open chain is equal to that in a periodic chain.

%
%

\subsection{\label{secB-1}Condition for continuum energy bands}
In a non-Hermitian system, the real-space Hamiltonian is given as
\begin{equation}
H=\sum_n\sum_{i=-N}^N\sum_{\mu,\nu=1}^qt_{i,\mu\nu}c_{n+i,\mu}^\dag c_{n,\nu},
\label{eqapp1b}
\end{equation}
where $t_{i,\mu\nu}$ is the hopping amplitude to the $i$th nearest unit cells, and it is not equal to $t_{-i,\nu\mu}^\ast$. In this Hamiltonian, $c_{n,\mu}$ is an annihilation operator of an electron with sublattice $\mu~(\mu=1,\dots,q)$ in the $n$th unit cell, and the electrons hop up to the $N$th nearest unit cells. Here, the solutions of the real-space eigenequation $H\ket{\psi}=E_{\rm OBC}\ket{\psi}$, where $\ket{\psi}=\left(\dots,\psi_{1,1},\dots,\psi_{1,q},\dots,\psi_{n,1},\dots,\psi_{n,q},\dots\right)^{\rm T}$, can be written as
\begin{equation}
\psi_{n,\mu}=\sum_{j=1}^{2M}\left(\beta_j\right)^n\phi_\mu^{\left(j\right)},~\left(\mu=1,\dots,q\right),
\label{eqapp2b}
\end{equation}
where $\beta=\beta_j$ is the solution of the characteristic equation
\begin{equation}
\det\left[{\cal H}\left(\beta\right)-E_{\rm OBC}\right]=0,
\label{eqapp3b}
\end{equation}
with the non-Bloch matrix
\begin{equation}
\left[{\cal H}\left(\beta\right)\right]_{\mu\nu}=\sum_{i=-N}^Nt_{i,\mu\nu}\beta^i,~\left(\mu,\nu=1,\dots,q\right).
\label{eqapp4b}
\end{equation}
We note that the characteristic equation is an algebraic equation for $\beta$ with an even degree $2M=2qN$ in general. Furthermore, we number the solutions of the characteristic equation so as to satisfy
\begin{equation}
\left|\beta_1\right|\leq\dots\leq\left|\beta_{2M}\right|.
\label{eqapp5b}
\end{equation}
Then the continuum energy bands in an infinite open chain can be obtained by the condition to the solutions of the characteristic equation, given by~\cite{Yokomizo2019}
\begin{equation}
\left|\beta_M\right|=\left|\beta_{M+1}\right|.
\label{eqapp6b}
\end{equation}
This is the main conclusion of the non-Bloch band theory. The trajectories of $\beta_M$ and $\beta_{M+1}$ give the generalized Brillouin zone, from which we can obtain the continuum energy bands in the limit of a large system size.

%
%

\subsection{\label{secB-2}Ladder model}
The ladder model in Eq.~(\ref{eq1}) with $\Delta\neq0$ is the case of $q=2$ and $N=1$, i.e., $M=2$ in Eq.~(\ref{eqapp6b}). Then because of Eq.~(\ref{eq10}), the condition for continuum energy bands can be explicitly written as
\begin{equation}
\left|\beta_2\right|=\left|\beta_3\right|=1.
\label{eqapp7b}
\end{equation}
Therefore, the generalized Brillouin zone is a unit circle, and the energy spectrum is described by the real Bloch wave number, i.e., $\beta={\rm e}^{ik},~k\in{\mathbb R}$. By substituting $\beta={\rm e}^{ik}$ into Eq.~(\ref{eq7}), we can get the energy spectrum in an open chain in the form of Eq.~(\ref{eq3}). Therefore, in the ladder model, the energy spectrum in an open chain matches with that in a periodic chain in the limit of a large system size.

%
%

\section{\label{secC}Derivation of Eq.~(\ref{eq11})}
In this appendix, we derive Eq.~(\ref{eq11}) from the characteristic equation (\ref{eq7}) and the boundary equation (\ref{eq8}). In the following, we focus on the case with the sufficiently large system size. In this case, the factors $\left(\beta_2\beta_4\right)^{L+1}$ and $\left(\beta_3\beta_4\right)^{L+1}$ are dominant on the left-hand side of Eq.~(\ref{eq8}). Hence, we can approximate Eq.~(\ref{eq8}) by
\begin{equation}
\left(\beta_2\right)^{2L+2}\simeq\frac{X_{1,2}X_{3,4}}{X_{1,3}X_{2,4}}
\label{eqapp1c}
\end{equation}
by using Eq.~(\ref{eq10}).

Next, we give perturbative solutions of Eq.~(\ref{eq7}) up to the second order of the coupling parameter $\Delta$. They can be explicitly written as
\begin{eqnarray}
\left\{ \begin{array}{l}
\beta_1\simeq x_-^{\left(1\right)}+y_-^{\left(1\right)}\Delta^2, \vspace{5pt}\\
\beta_2\simeq x_+^{\left(1\right)}+y_+^{\left(1\right)}\Delta^2, \vspace{5pt}\\
\beta_3\simeq x_-^{\left(2\right)}+y_-^{\left(2\right)}\Delta^2, \vspace{5pt}\\
\beta_4\simeq x_+^{\left(2\right)}+y_+^{\left(2\right)}\Delta^2,
\end{array}\right.
\label{eqapp2c}
\end{eqnarray}
where
\begin{eqnarray}
\left\{ \begin{array}{l}
\displaystyle x_\pm^{\left(1\right)}=\frac{1}{2t_L}\left(E_{\rm OBC}\pm\delta\right), \vspace{5pt}\\
\displaystyle x_\pm^{\left(2\right)}=\frac{1}{2t_R}\left(E_{\rm OBC}\pm\delta\right), \vspace{5pt}\\
\displaystyle y_\pm^{\left(1\right)}=\frac{\pm\left(E_{\rm OBC}^2-2t_Lt_R\pm\delta E_{\rm OBC}\right)}{\left(t_L-t_R\right)\delta\left[2t_L\left(t_L+t_R\right)-E_{\rm OBC}^2\mp\delta E_{\rm OBC}\right]}, \vspace{5pt}\\
\displaystyle y_\pm^{\left(2\right)}=\frac{\pm\left(E_{\rm OBC}^2-2t_Lt_R\pm\delta E_{\rm OBC}\right)}{\left(t_L-t_R\right)\delta\left[-2t_R\left(t_L+t_R\right)+E_{\rm OBC}^2\pm\delta E_{\rm OBC}\right]},
\end{array}\right. \nonumber\\
\label{eqapp3c}
\end{eqnarray}
and
\begin{equation}
\delta=\sqrt{E_{\rm OBC}^2-4t_Lt_R}.
\label{eqapp4c}
\end{equation}
From Eq.~(\ref{eqapp2c}), we can calculate the right-hand side of Eq.~(\ref{eqapp1c}). In fact, the factors included in the right-hand side of this equation can be given by
\begin{eqnarray}
\left\{ \begin{array}{l}
\displaystyle X_{1,2}=\frac{t_L+t_R}{t_L-t_R}\frac{\delta}{\left(t_L+t_R\right)^2-E_{\rm OBC}^2}\Delta^2+{\cal O}\left(\Delta^4\right), \vspace{5pt}\\
\displaystyle X_{3,4}=\left(\frac{t_L}{t_R}-\frac{t_R}{t_L}\right)\delta+{\cal O}\left(\Delta^2\right), \vspace{5pt}\\
\displaystyle X_{1,3}=\left(t_L-t_R\right)\left(\frac{E_{\rm OBC}-\delta}{2t_R}-\frac{2t_R}{E_{\rm OBC}-\delta}\right)+{\cal O}\left(\Delta^2\right), \vspace{5pt}\\
\displaystyle X_{2,4}=\left(t_L-t_R\right)\left(\frac{E_{\rm OBC}+\delta}{2t_R}-\frac{2t_R}{E_{\rm OBC}+\delta}\right)+{\cal O}\left(\Delta^2\right).
\end{array}\right. \nonumber\\
\label{eqapp5c}
\end{eqnarray}
As a result, we can get Eq.~(\ref{eq11}) by substituting Eq.~(\ref{eqapp5c}) into Eq.~(\ref{eqapp1c}) and approximating the equation up to the second order of $\Delta$.

%
%

\section{\label{secD}Derivation of Eq.~(\ref{eq19})}
In this appendix, we briefly explain the derivation of Eq.~(\ref{eq19}). In the following, we focus on a non-Hermitian system in Eq.~(\ref{eqapp1b}) with an open boundary condition with the system size $L$. In the following, we assume that the system has some symmetries which reverse the direction of the chain, such as inversion symmetry. We note that, in this case, the characteristic equation (\ref{eqapp3b}) is a reciprocal equation, and its solutions satisfy $\beta_j=1/\beta_{2M+1-j}~(j=1,\dots,M)$. Now we have $2M$ boundary conditions to the eigenstates in Eq.~(\ref{eqapp2b}). Although Eq.~(\ref{eqapp2b}) has the $2qM$ unknown combination coefficients, from the real-space eigenequation $H\ket{\psi}=E_{\rm OBC}\ket{\psi}$, we can reduce the $2qM$ coefficients to the $2M$ coefficients, e.g., $\phi_1^{\left(j\right)}~(j=1,\dots,2M)$. As a result, the $2M$ boundary conditions to the $2M$ coefficients can be obtained from the left and right ends of an open chain. Finally, we have the condition so that the coefficients $\phi_1^{\left(j\right)}~(j=1,\dots,2M)$ take nonzero values, written as
\begin{equation}
\sum_{P,Q}F\left(\beta_{i\in P},\beta_{j\in Q},E_{\rm OBC}\right)\prod_{k\in P}\left(\beta_k\right)^L=0,
\label{eqapp1d}
\end{equation}
where the sets $P$ and $Q$ are two disjoint subsets of the set $\{1,\dots,2M\}$ with $M$ elements, and the sets of the solutions of Eq.~(\ref{eqapp3b}) determine the values of the function $F$. In the limit of a large system size, we can reduce Eq.~(\ref{eqapp1d}) to
\begin{equation}
\left(\frac{\beta_M}{\beta_{M+1}}\right)^L=-\frac{F\left(\beta_{i\in P_0},\beta_{j\in Q_0},E_{\rm OBC}\right)}{F\left(\beta_{i\in P_1},\beta_{j\in Q_1},E_{\rm OBC}\right)},
\label{eqapp2d}
\end{equation}
with $P_0=\{M+1,\dots,2M\},Q_0=\{1,\dots,M\},P_1=\{M,M+2,\dots,2M\}$, and $Q_1=\{1,\dots,M-1,M+1\}$. This is because Eq.~(\ref{eqapp1d}) has two leading terms proportional to $\left(\beta_M\beta_{M+2}\cdots\beta_{2M}\right)^L$ and $\left(\beta_{M+1}\beta_{M+2}\cdots\beta_{2M}\right)^L$. Then, because of $\beta_M=1/\beta_{M+1}$, we can write Eq.~(\ref{eqapp2d}) in the form of Eq.~(\ref{eq19}).

%

\providecommand{\noopsort}[1]{}\providecommand{\singleletter}[1]{#1}%
\end{document}